\documentclass[aps,showpacs,manuscript,12pt]{revtex4}
\usepackage{amssymb}
\usepackage{amsmath}
\usepackage{graphicx}

\begin{document}
\title{\textbf{Strong solutions of the incompressible Navier-Stokes equations in
external domains: local existence and uniqueness}}
\author{M. Tessarotto$^{a,b}$ and C. Cremaschini$^{c}$}
\affiliation{\ $^{a}$Department of Mathematics and Informatics,
University of Trieste, Italy\\ $^{b}$Consortium of
Magneto-fluid-dynamics, University of Trieste, Italy \\
$^{c}$International School for Advanced Studies, SISSA, Trieste,
Italy }

\begin{abstract}
In this paper the problem of strong solvability of the
incompressible Navier-Stokes equations (INSE) is revisited, with
the goal of determining the minimal assumptions for the validity
of a local existence and uniqueness theorem for the Navier-Stokes
fluid fields (solutions of INSE). Emphasis is placed on fluid
fields which, together with suitable derivatives, do not
necessarily decay at infinity and hence do not belong to Sobolev
spaces. For this purpose a novel approach based on a so-called
inverse kinetic theory, recently developed by Tessarotto and
Ellero, is adopted. This requires the construction of a suitable
kinetic equation, advancing in time a suitably smooth kinetic
distribution function and providing exactly, as its moment
equations, the complete set of fluid equations. In turn, by proper
definition of the kinetic equation, this permits the introduction
of the so-called \textit{Navier-Stokes dynamical system}, i.e.,
the dynamical system which advances in time self-consistently the
Navier-Stokes fluid fields. Investigation of the properties of
this dynamical system is crucial for the establishment of an
existence and uniqueness theorem for strong solutions of INSE. The
new theorem applies both to bounded and unbounded domains and in
the presence of generalized boundaries, represented by surfaces,
curves or even sets of isolated points. In particular, for
unbounded domains, solutions are considered, which do not
necessarily vanish at infinity. Basic consequences for the
functional setting of classical solutions are analyzed.
\keywords{Navier-Stokes equations \and Kinetic theory \and
Dynamical systems }
\end{abstract}

\pacs{47.10.ad,47.10.Fg,47.10.A-}
\date{\today }
\maketitle

{\bf \noindent MSC Primary 76D03 Secondary 76D06}



\section{Introduction}

\label{Intro}

The goal of this paper is to investigate a well-known theoretical
issue of fluid dynamics. This is concerned with the existence and
smoothness of solutions of the fluid equations which characterize
an incompressible Newtonian fluid, i.e., the so-called
\emph{incompressible Navier-Stokes} \emph{equations}
(\emph{INSE}). In particular, we will be interested in solutions
which are physically realizable, i.e., can be identified with
\emph{physical observables}. In the theory of continua these are
necessarily described by strong solutions. Several aspects of the
mathematical theory of these fluid equations remain unsolved. This
occurs, in particular, in the case of:

\begin{itemize}
\item \textit{3D} \textit{external domains;}

\item \textit{forcing;}

\item \textit{solutions which do not necessarily decay at infinity in
configuration space.}
\end{itemize}

A related problem of mathematical research involves the search of the \emph{%
classical dynamical system} - here denoted as \emph{Navier-Stokes
(NS) dynamical system} - which uniquely advances in time the
related fluid fields
that characterize the Newtonian fluid, i.e., the fluid velocity $\mathbf{V}(%
\mathbf{r,}t)$ and the fluid pressure$\ p(\mathbf{r,}t)$. In other
words, if the NS dynamical system actually exists, it will permit
to cast the complete set of fluid equations in terms of an
equivalent (and possibly infinite) set of ordinary differential
equations which define the dynamical system itself. From the
mathematical viewpoint the reason why such a dynamical system is
so important is that, if a theorem of existence and uniqueness can
be established for the set of ordinary differential equations
which determines the dynamical system, than it is obvious that it
will imply the same conclusion also for INSE. For contemporary
science the determination of such a dynamical system represents
not merely an unsolved intellectual challenge, but a fundamental
prerequisite for the proper formulation of all physical theories
which are based on the description of these fluids. These involve,
for example, the understanding of the related phase-space
Lagrangian fluid descriptions and the consistent formulation of
turbulence theory and of the related approximate statistical
descriptions (i.e., obtained introducing appropriate stochastic
models able to reproduce phenomenological data), both essential in
fluid dynamics and in applied sciences.

Surprisingly, until recently (see Tessarotto and Ellero, 2000-2005 \cite%
{Ellero2000,Ellero2004,Ellero2005}), the problem has remained
unsolved. Its
solution, discussed in detail in Refs. \cite%
{Ellero2005,Tessarotto2006,Tessarotto2007}, is based on the
construction of an \emph{inverse kinetic theory} (\emph{IKT}) for
the incompressible NS equations, i.e., a suitable phase-space
description which provides the complete set of fluid equations in
terms of an appropriate inverse kinetic equation. Such an
equation, in particular, can be constructed in such a way to
define uniquely a classical dynamical system which advances in
time the fluid fields, i.e., so that they result - by construction
- solutions of the same fluid equations. For the extension of the
theory to the treatment of
incompressible thermofluids and quantum fluids see Refs.\cite%
{Tessarotto2008a,Tessarotto2008b,Piero,Piero2}.

In this paper we intend to point out basic consequences which can
be drawn from IKT as far as the problem of existence and
smoothness of solutions of INSE is concerned.

\subsection{Strong solutions of the INSE problem}

\label{sec:1}

For definiteness, {we shall assume that the relevant fluids }$\left\{ \rho ,%
\mathbf{V},p\right\} ,${\ i.e., respectively the mass density,
fluid velocity and pressure describing the fluid, are defined
point-wise and suitably smooth in an appropriate domain }$\Omega
\times $$I${. Here $\Omega
$ (\emph{fluid domain}) is an open connected subset of the Euclidean space }$%
\mathbb{R}^{3}${, with boundary} $\partial \Omega $ and{\ closure }$%
\overline{\Omega }.$ $\overline{\Omega }$ is defined as the set
where the mass density is a constant $\rho \equiv \rho _{o}>0${;
moreover, }$I${\ is
either a time interval $I=$}$\left] {t_{0},t_{1}}\right[ $ ({with closure }$%
\overline{{I}}${$=[t_{0},t_{1}]$)$,$ or the real axis }$%
\mathbb{R}
$. We assume that the fluid fields are continuous in $\overline{\Omega }%
\times \overline{{I}},$ fulfill suitable initial and boundary
conditions respectively at $t=t_{o}$ and on the boundary $\partial
\Omega ,$ while in the open set $\Omega \times ${$I$ they satisfy
the fluid equations:} {\
\begin{eqnarray}
&&\left. \frac{D}{Dt}\rho +\rho \nabla \cdot \mathbf{V}=0,\right.
\label{1}
\\
&&\left. \rho \frac{D}{Dt}\mathbf{V}+\nabla p-\mathbf{f}-\mu \nabla ^{2}%
\mathbf{V}=\mathbf{0},\right.  \label{2} \\
&&\left. \nabla \cdot \mathbf{V}=0,\right.  \label{3} \\
&&\left. \rho =\rho _{o},\right.  \label{5a}
\end{eqnarray}%
which are subject to the inequalities}%
\begin{eqnarray}
&&\left. p>0,\right.  \label{5aa} \\
&&\left. \rho >0.\right.  \label{6aa}
\end{eqnarray}%
Here{\ the notation is standard. Thus, }$\rho _{o}$ and $\mu >0$
are
respectively the constant mass density and fluid viscosity and{\ }$\mathbf{f}%
(\mathbf{r,}t)$ the volume force density. \ Moreover,
$\frac{D}{Dt}\equiv \frac{\partial }{\partial
t}\mathbf{V}+\mathbf{V}\cdot \nabla $ is the convective
derivative, {Eq.(\ref{1}) denotes the so-called continuity
equation, while Eqs.(\ref{2}), (\ref{3}) and (\ref{5a}) are
respectively the \emph{Navier-Stokes }}(NS), {\emph{isochoricity
}} and {\emph{incompressibility}} equations. Finally, the inequalities (\ref{5aa}%
) and (\ref{6aa}) {represent the so-called \emph{physical
realizability
conditions} of the fluid, which must be prescribed in order that }$p$ and $%
\rho $ are \emph{physical observables.}

The complete set of equations (\ref{1})-(\ref{5a}) - subject to
the
inequalities (\ref{5aa}) and (\ref{6aa}) -{\ are denoted as \emph{%
incompressible Navier-Stokes equations} (\emph{INSE}) and its
solutions \emph{NS fluid fields}. In the following }it is assumed
that: 1){\ }$\left\{ \rho ,\mathbf{V},p\right\} $ are strong
solutions of the INSE problem which are suitably smooth so that
{Eqs.(\ref{5}) and (\ref{6}) are identically satisfied in the
whole set }$\Omega \times I;$ 2) {the solutions\ }of INSE are
assumed to {satisfy suitable initial-(Dirichlet-)boundary value
problem\ (\emph{INSE problem}). } These are defined as follows:

\noindent A) \textit{fluid
initial conditions}: the initial conditions for the fluid fields $A(\mathbf{r%
},t_{o})$ $\equiv \left\{ \rho ,\mathbf{V,}p\right\} _{\mathbf{(r,}t_{o}%
\mathbf{)}}$ are defined imposing
\begin{equation}
A\mathbf{(r,}t_{o}\mathbf{)=A}_{o}(\mathbf{r}),  \label{initial
contdion -1}
\end{equation}%
where $t_{o}\in I,$ and $\mathbf{A}_{o}(\mathbf{r})\equiv \left\{ \rho _{o},%
\mathbf{V}_{o}(\mathbf{r}),p_{o}(\mathbf{r})\right\} $ satisfy
respectively the condition of isochoricity and the Poisson
equation:
\begin{equation}
\nabla \cdot \mathbf{V}_{o}=0,  \label{initial condition -2}
\end{equation}%
\begin{equation}
\nabla ^{2}p_{o}=-\nabla \cdot \left[ \mathbf{V}_{o}\mathbf{\cdot \nabla V}%
_{o}\right] \mathbf{-\nabla \cdot f}(\mathbf{r,}t_{o}).
\label{initial condition -3}
\end{equation}%
B) \textit{fluid boundary conditions}: $\partial \Omega $ is
considered for
greater generality as a moving boundary. In particular, for all points $%
\mathbf{r}_{W}(t)\in \partial \Omega $ let us assume that their
velocity
\begin{equation}
\mathbf{V}_{w}(\mathbf{r}_{W}(t),t)\equiv
\frac{d}{dt}\mathbf{r}_{W}(t) \label{boundary velocity}
\end{equation}%
is a suitably smooth function of time defined in $I$. Then, the
boundary conditions are obtained, in both cases, imposing the
Dirichlet boundary conditions
\begin{equation}
\lim_{\mathbf{r\in \Omega \rightarrow r}_{W}\in \partial \Omega }A(\mathbf{r,%
}t)=A_{W}(\mathbf{r}_{W}\mathbf{,}t),  \label{boundary condition}
\end{equation}%
where $A_{W}(\mathbf{r}_{W}\mathbf{,}t)\equiv \left\{ \rho _{o},\mathbf{V}%
_{W}\mathbf{,}p_{W}\right\}
_{\mathbf{(r}_{W}\mathbf{,}t\mathbf{)}}.$

\bigskip

In particular, we remark that:

\begin{itemize}
\item by taking the divergence of the NS equation (\ref{2}) and respectively
its scalar product with $\mathbf{V,}$ it follows the \emph{Poisson
equation for the fluid pressure} $p$
\begin{equation}
\nabla ^{2}p=-\rho _{o}\nabla \cdot \left( \mathbf{V}\cdot \nabla \mathbf{V}%
\right) +\nabla \cdot \mathbf{f}  \label{5}
\end{equation}%
and the \emph{energy equation}%
\begin{equation}
\rho \frac{D}{Dt}\frac{V^{2}}{2}+\mathbf{V}\cdot \left[ \nabla p-\mathbf{f}%
-\mu \nabla ^{2}\mathbf{V}\right] =0.  \label{6}
\end{equation}

\item since conventionally the domain of vacuum is defined as the subdomain
(of $\mathbb{R}^{3}$) in which both $p$ and $\rho $ vanish
identically, the inequalities (\ref{5aa}) and (\ref{6aa}) provide
the physical requirements in order that the domain $\Omega $ is
non-empty;

\item the boundary of the fluid domain, $\partial \Omega $, may generally
include so-called \emph{free boundaries} where the fluid pressure $p(\mathbf{%
r,}t)$ locally vanishes;

\item in case $\Omega $ is unbounded and $\partial \Omega $ (or a subset of $%
\partial \Omega $) is an improper surface of $\mathbb{R}^{3},$ the fluid
fields $A(\mathbf{r,}t)$ need not decay necessarily at infinity.
This means that $A(\mathbf{r}_{W}\mathbf{,}t)$ may not vanish at
infinity. As a consequence, in such a case it follows that
$\mathbf{V(r,}t\mathbf{)}$ shall not be required to be
$L^{2}(\Omega ).$
\end{itemize}

\subsection{Motivations and historical background}

\label{sec:2}

The modern status of mathematical theory of NS equations is
largely due to
the pioneering work of Leray in the years 1933-34 \cite%
{Leray1933,Leray1934a,Leray 1934b} and Hopf\ in 1950-51
\cite{Hopf1950/51} who reformulated the NS PDE's in terms of a
suitable set of integral equations and introduced the concept of
\textit{weak solutions} for NS equations. These solutions have
played, since their introduction, a major role in the mathematical
research dealing with NS equations. Indeed, they are the only
solutions which, so far, have been proven to exist for all times
and without restrictions on the initial data, apart the
requirement of a suitable functional setting for the fluid fields
(for a review see Galdi,\ 2002 \cite{Galdi2002}). However, the
fundamental problem of a unique global weak solution and,
conversely, the possibility that uniqueness holds only locally due
to the appearance of a local "bifurcation" phenomenon (the
so-called Leray conjecture \cite{Temam1979}) still remain open
issues.

On the other hand, the distinction between strong (or
\textit{classical}) and weak solutions has important relevance
also for the modelling of real fluids in the framework of
continuous mechanics. It should be pointed out that the existence
of weak solutions, rather than the classical ones, for NS
equations may be also potentially an indication of the failure of
the mathematical model based on the incompressible Navier-Stokes
equations. In fact, it may involve the violation of the subsidiary
conditions indicated above, in particular since a weak solution is
generally
not defined everywhere in $\Omega \times I$ (Doering and Gibbon, 1997 \cite%
{Doering1997}).

An important feature which characterizes weak solutions is the way
in which they are usually manufactured, i.e., their existence and
uniqueness is established. This is obtained replacing the NS PDE's
by an infinite set of ODE's and by constructing explicitly their
solutions, which furnish a sequence of successive approximations,
the so-called \textit{Galerkin approximates} (\cite{Doering1997}).
The existence and uniqueness theorem for weak solutions of NS\
equations is achieved by demonstrating that the Galerkin
approximates converge in a suitably weak sense to the solutions of
the original PDE's (which is another way of justifying the name
given to these solutions).

In its traditional approach \cite{Leray
1934b,Hopf1950/51,Lions1959,Serrin1963,Ladyzhenskaya1969,Kato1964,Kato1965,Kato}
the treatment of the existence and uniqueness problem in the local
sense, i.e., for a finite time interval $I,$ requires a suitable
functional setting which depends, in particular, on the choice of
the boundary conditions.\ A necessary condition for the fluid
fields to be strong solutions of INSE is that they are one-sided
continuous on boundaries which separate the configuration space.
Moreover, in the open domain $\Omega ,$ since all fluid
equations [i.e., Eqs.(\ref{1})-(\ref{5a}) as well as (\ref{5}) and (\ref{6}%
)] must be defined by functions which are at least continuous, i.e., $%
C^{(0,0)}(\Omega \times I),$ the fluid fields and the volume force density $%
\mathbf{f}\mathbb{(}\mathbf{r,}t)$ which satisfy INSE must
evidently have the \emph{native functional setting (NF setting)}.
These conditions involve both the fluid fields $\left\{ \rho
,\mathbf{V},p\right\} $ and the volume force
$\mathbf{f}(\mathbf{r,}t),$ which are required to satisfy at least

\begin{equation}
\left\{
\begin{array}{c}
\left\{ \rho ,\mathbf{V},p\right\} \in C^{(0)}(\overline{\Omega
}\times I),
\\
\mathbf{V}(\mathbf{r,}t)\mathbf{\in }C^{(3,1)}(\Omega \times I), \\
p(\mathbf{r,}t)\mathbf{\in }C^{(2,0)}(\Omega \times I), \\
\mathbf{f}(\mathbf{r,}t)\mathbf{\in }C^{(1,0)}(\Omega \times I),%
\end{array}%
\right.  \label{NSF functional setting-1}
\end{equation}%
as well as their initial and boundary conditions, which imply also%
\begin{equation}
\left\{
\begin{array}{c}
\mathbf{V}_{o}(\mathbf{r})\mathbf{\in }C^{(3)}(\Omega ), \\
p_{o}(\mathbf{r})\mathbf{\in }C^{(2)}(\Omega ), \\
\mathbf{f}(\mathbf{r,}t_{o})\mathbf{\in }C^{(1)}(\Omega ),%
\end{array}%
\right.  \label{NSF functional setting-2}
\end{equation}%
and%
\begin{equation}
\left\{
\begin{array}{c}
\mathbf{V}_{W}(\mathbf{r}_{W}(t),t)\mathbf{\in }C^{(1)}(I), \\
p_{W}(\mathbf{r}_{W}(t),t)\mathbf{\in }C^{(1)}(\Omega ).%
\end{array}%
\right.  \label{NSF functional setting 3}
\end{equation}

Indeed, in validity of (\ref{NSF functional setting-1}) [and
(\ref{NSF functional setting-2}),(\ref{NSF functional setting
3})], by invoking
equations (\ref{2})-(\ref{3}) and imposing in the whole set $\overline{%
\Omega }$ the initial condition
\begin{equation}
\rho (\mathbf{r},t_{o})=\rho _{o}>0,  \label{7}
\end{equation}%
it is obvious that Poisson equation Eq.(\ref{5}) implies the
incompressibility and the isochoricity conditions (\ref{3}) and
(\ref{5a}),
as well as the energy equation (\ref{6}) and vice versa. Hence, imposing Eq.(%
\ref{7}), INSE can be reduced to the equivalent set given by Eqs.(\ref{1})-(%
\ref{5a}) and the physical realizability conditions
(\ref{5aa}),(\ref{6aa}).
Here $\overline{\Omega }=\Omega \cup \partial \Omega $ is the closure of $%
\Omega $ and $C^{(i,j)}(\Omega \times I)\equiv C^{(i)}(\Omega
)\cap C^{(j)}(I).$

The customary approach to weak solutions is, instead, based on the
requirement that the fluid fields, together with the data, belong
to appropriate Sobolev spaces \cite{Ladyzhenskaya1969,Lions,Kato1964,Temam1983}%
, endowed with a suitable scalar product, usually defined by the
integral on the configuration space $\Omega $
\begin{equation}
\left(
\mathbf{V}(\mathbf{r,}t),\mathbf{V}_{1}(\mathbf{r,}t)\right)
\equiv \int_{\Omega }d\mathbf{rV}(\mathbf{r,}t)\cdot
\mathbf{V}_{1}(\mathbf{r,}t),
\end{equation}%
being $\mathbf{V}(\mathbf{r,}t),\mathbf{V}_{1}(\mathbf{r,}t)$ two
independent velocity fields, and with norm
\begin{equation}
\left\Vert \mathbf{V}(\mathbf{r,}t)\right\Vert _{1}^{2}\equiv \int_{\Omega }d%
\mathbf{r}\sum\limits_{\alpha =0,1}\left( D^{\alpha }\mathbf{V,}D^{\alpha }%
\mathbf{V}\right) .
\end{equation}%
The latter definition involves, besides the velocity field itself $\mathbf{V=%
}$ $D^{0}\mathbf{V}$ , its gradient $\nabla
\mathbf{V}(\mathbf{r,}t)\equiv D^{1}\mathbf{V.}$ If the domain
$\Omega $ is bounded and the fluid fields belong to the NF setting
both the scalar product and the norm can be trivially defined. \
However, this is not generally true in the case of an
external (i.e., unbounded) domain $\Omega ,$ unless the velocity field $%
\mathbf{V}(\mathbf{r,}t)$ vanishes for $\left\vert
\mathbf{r}\right\vert \rightarrow \infty $ . This is usually
denoted \emph{Sobolev-space functional setting (SB setting)}.

\subsection{Open problems}

\label{sec:2b}

Referring to the $3D$ case the theory of existence and uniqueness
of solutions is not yet complete in several aspects.

In particular regarding existence, a major open question is
whether strong solutions exist for all times (\emph{global strong
solutions}), i.e., the fluid fields $A(\mathbf{r,}t)$ remain at
all times in the same (NF) functional class, or cease to exist in
a proper sense, i.e., they develop spontaneously singularities
which violate, at least locally, the Navier-Stokes equations. \ In
both cases the answer would be of useful: in the first case to
establish a global existence and uniqueness theorem with
far-reaching consequences; in the second, to understand the nature
of "singularities" or "irregularities" (i.e., for example,
discontinuities arising in suitably higher-order derivatives of
the fluid fields or violations of suitable bounds in appropriate
functional spaces) possibly produced by the spontaneous evolution
and decay of turbulence in a finite or infinite time interval
\cite{Temam1979}.

Another issue of fundamental importance in its own right is the
problem of \emph{uniqueness of strong solutions}, which are
assumed to exist and to belong to a suitable functional class (for
example NF or SB). Several authors have investigated\ the role of
functional settings and supplementary regularity assumptions to be
satisfied by the fluid fields and/or the data. In this regard, of
fundamental importance is, in particular, the result obtained by \
Ladyzhenskaya \cite{Ladyzhenskaya1969}, who was able to prove the
uniqueness of the fluid fields $A(\mathbf{r,}t\mathbf{),}$ assumed
to exist as strong solutions belonging to the SB setting, by
imposing the additional regularity constraint\ that the velocity
field remains bounded in $L^{4}(\Omega ).$ An analogous result was
obtained later for strong
solutions in $L^{p}(\Omega ),$ with $p\geq 2,$ (Kato and Ponce, 1984 \cite%
{Kato}; Sohr and von Wahl, 1986 \cite{von Wahl1986}; Deuring and
von Wahl, 1995 \cite{Von Wahl1995} ) and has been addressed by
several authors also for weak solutions (see, for example,
Iftimie, 1999 \cite{Iftimie1999}, Koch and Tataru, 2001
\cite{Koch2001}, Galdi, 2002 \cite{Galdi2002})$.$ However, these
results fail, in general, in the case of an external domain
$\Omega ,$ for the same reason indicated above. Therefore, a
fundamental problem, as yet unanswered, both for weak and strong
solutions, is the search of the minimal functional setting for the
existence of \emph{local} and respectively \emph{global
solutions}\textit{\ }in exterior domains (Galdi
and Maremonti, 1986 \cite{Galdi1986}) and which do not decay at infinity in%
\textit{\ }$\Omega .$\textit{\ }Additional issues concern both \emph{%
uniqueness} (J.Kato, 2003 \cite{Jun-Kato2003}) and
\emph{existence} (Giga et al.,1999 \cite{Giga1999}; Giga et al.
\cite{Giga2001}).

\subsection{Goal of the paper}

\label{sec:3}

Goal of the paper is to address the problem of strong solvability
of the incompressible Navier-Stokes equations (\textit{INSE}) in
$\Omega \times I.$
$\Omega $ is here identified with a subset of the Euclidean space $%
\mathbb{R}
^{3}$, formed at most by a finite number of disjoint, open and
connected
subsets $\Omega _{i}$ of non-vanishing, and possibly non-finite measure (%
\emph{fluid subdomains}); as a particular case, $\Omega $ ($\Omega
_{i})$
can also be identified with an unbounded subdomain of $%
\mathbb{R}
^{3}$ (\emph{external domain}). Moreover, $I$ is subset of $R,$
assumed
either bounded or unbounded. In the following we intend to seek the \emph{%
minimal functional setting} for the well-posedness of local strong
solutions which, together with suitable derivatives of the fluid
fields, in the case in which $\Omega $ is an external domain, do
not necessarily decay at infinity in $\Omega $ (\emph{nondecaying
strong solutions}) and hence do not belong to Sobolev spaces.
\emph{More precisely, this involves the search of
appropriate minimal functional classes for the NS fluid fields }$V(\mathbf{r}%
,t),$ $p(\mathbf{r},t),$\emph{\ assuring the validity of a local
existence and uniqueness theorem for the initial-boundary value
problem for NS equations.}

The treatment here developed is based on the IKT\textit{\ }earlier
developed by Tessarotto and Ellero
\cite{Ellero2005,Tessarotto2006,Tessarotto2007}, which relies on
the construction of a suitable kinetic equation and providing
exactly, as its moments equations, the required set of fluid
equations. The IKT here adopted is shown to provide the basis for
a local existence and uniqueness theorem for strong solutions of
the NS equations.

Key features of the present treatment are as follows:

1) The fluid domain $\Omega $ is a three-dimensional subset of $%
\mathbb{R}
^{3}$ which is not necessarily bounded, i.e., it can be an
external domain.

2) In the case in which $\Omega $ is unbounded, the fluid fields,
and in particular the fluid velocity $\mathbf{V}(\mathbf{r,}t),$
are not required to be $L^{p}(\Omega )$ in the fluid domain
$\Omega $, with $p\geq 2.$ Hence,
the fluid fields $A(\mathbf{r},t)=\left\{ \rho \mathbf{,V,}p\right\} \mathbf{%
,}$ together with their derivatives $\frac{\partial }{\partial t}A(\mathbf{r}%
,t),\nabla A(\mathbf{r},t),\nabla ^{2}A(\mathbf{r},t)$ and the
volume force density $\mathbf{f}(\mathbf{r,}t),$ are permitted to
be nondecaying, i.e.,
to admit non-vanishing asymptotic values in $\Omega $. Thus, denoting $%
\widehat{e}_{\mathbf{r}}$ the unit vector $\widehat{e}_{\mathbf{r}}=\mathbf{%
r/}r$ and $B\equiv \left\{ A(\widehat{e}_{\mathbf{r}},t),\frac{\partial }{%
\partial t}A,\mathbf{\nabla }A,\mathbf{\nabla }^{2}A,\mathbf{f}\right\} ,$
it must be such that%
\begin{equation}
\left\{
\begin{array}{c}
\lim_{\mathbf{r\in \Omega ,}\left\vert \mathbf{r}\right\vert
\rightarrow
\infty }B(\mathbf{r,}t)\mathbf{=}B_{_{\infty }}(\widehat{e}_{\mathbf{r}},t)%
\mathbf{,} \\
\left. 0\leq \left\vert B_{_{\infty }}(\widehat{e}_{\mathbf{r}%
},t)\right\vert <\infty \right. ,%
\end{array}%
\right.
\end{equation}%
where the limits $B_{\infty }\equiv \left\{ A_{_{\infty }}(\widehat{e}_{%
\mathbf{r}},t),\frac{\partial }{\partial t}A_{\infty },\mathbf{\nabla }%
A_{\infty },\mathbf{\nabla }^{2}A_{\infty },\mathbf{f}_{\infty
}\right\} $ are allowed to be non-zero.

3) The fluid pressure $p(\mathbf{r,}t)$ is assumed non negative in
the closure domain $\overline{\Omega },$ i.e., $p\geq 0,$ thus
permitting the
existence of \textit{vacuum regions}, i.e., subdomains of $%
\mathbb{R}
^{3}$in which $\rho $ and $p$ vanish identically, and corresponding \textit{%
free boundaries}, i.e., the parts of (the border of)
$\overline{\Omega }$ on which the fluid pressure vanishes
identically.

4) The boundary set $\partial \Omega $ may generally be formed by\
piece-wise regular surfaces, curves, as well as isolated points of $\overline{%
\Omega }\subseteq
\mathbb{R}
^{3}.$

5) In validity of assumptions 1)-3), a local existence and
uniqueness theorem for strong solutions of the initial-boundary
value problem for INSE
is found by assuming that the fluid fields $\left\{ \rho ,\mathbf{V,}%
p\right\} $ and the volume force density acting on the fluid in
$\Omega \times I$ \ satisfy stronger requirements than those
provided by the NF
setting and (\ref{5aa},\ref{6aa})\ which are given by the following assumptions%
\begin{equation}
\left\{
\begin{array}{c}
\left\{ \rho ,\mathbf{V},p,\underline{\underline{\mathbf{\Pi }}},\mathbf{Q,f}%
(\mathbf{r,}t)\right\} \in C^{(0)}(\overline{\Omega }\times I), \\
\mathbf{V}(\mathbf{r,}t)\mathbf{\in }C^{(3,2)}(\Omega \times I), \\
\underline{\underline{\mathbf{\Pi }}},\mathbf{Q,}p(\mathbf{r,}t)\mathbf{\in }%
C^{(2,2)}(\Omega \times I), \\
\mathbf{f}(\mathbf{r,}t)\mathbf{\in }C^{(2,0)}(\Omega \times I),%
\end{array}%
\right.  \label{minimal functional setting N-1}
\end{equation}%
while
\begin{equation}
\left\{
\begin{array}{c}
\mathbf{V}_{o}(\mathbf{r})\mathbf{\in }C^{(3)}(\Omega ), \\
\underline{\underline{\mathbf{\Pi }}}_{o},\mathbf{Q}_{o}\mathbf{,}p_{o}(%
\mathbf{r})\mathbf{\in }C^{(2)}(\Omega ), \\
\mathbf{f}(\mathbf{r,}t_{o})\mathbf{\in }C^{(2)}(\Omega ).%
\end{array}%
\right.  \label{minimal functional setting N-2}
\end{equation}%
Moreover, the velocity $\mathbf{V}_{w}(\mathbf{r}_{W}(t),t)$ of each point, $%
\mathbf{r}_{W}(t),$ of the boundary $\partial \Omega $ is assumed
at least of class
\begin{equation}
\mathbf{V}_{w}(\mathbf{r}_{W}(t),t)\in C^{(2)}(I). \label{minimal
functional setting N-2b}
\end{equation}

\noindent Here $\underline{\underline{\mathbf{\Pi }}}(\mathbf{r,}t),\mathbf{Q}(\mathbf{%
r,}t)$ and $\underline{\underline{\mathbf{\Pi }}}_{o}(\mathbf{r}),\mathbf{Q}%
_{o}(\mathbf{r})$ denote auxiliary fluid fields to be suitably
defined. Furthermore, \ in the case in which $\Omega $ is
unbounded, the fluid fields
$A(\mathbf{r},t)=\left\{ \rho \mathbf{,V,}p,\underline{\underline{\mathbf{%
\Pi }}},\mathbf{Q}\right\} $ and the volume force density $\mathbf{f}(%
\mathbf{r,}t),$ together with their derivatives $\frac{\partial }{\partial t}%
A(\mathbf{r},t),\nabla A(\mathbf{r},t),\nabla
^{2}A(\mathbf{r},t),$ are permitted to be nondecaying, i.e., to
admit non-vanishing asymptotic values
in $\Omega $. Thus, denoting $\widehat{e}_{\mathbf{r}}$ the unit vector $%
\widehat{e}_{\mathbf{r}}=\mathbf{r/}r$ and $B\equiv \left\{ A(\widehat{e}_{%
\mathbf{r}},t),\frac{\partial }{\partial t}A,\mathbf{\nabla }A,\mathbf{%
\nabla }^{2}A,\mathbf{f}\right\} ,$ it must be%
\begin{equation}
\left\{
\begin{array}{c}
\lim_{\mathbf{r\in \Omega ,}\left\vert \mathbf{r}\right\vert
\rightarrow
\infty }B(\mathbf{r,}t)\mathbf{=}B_{_{\infty }}(\widehat{e}_{\mathbf{r}},t)%
\mathbf{,} \\
\left. 0\leq \left\vert B_{_{\infty }}(\widehat{e}_{\mathbf{r}%
},t)\right\vert <\infty \right. ,%
\end{array}%
\right.  \label{minimal functional setting N-3}
\end{equation}%
where the limits $B_{\infty }\equiv \left\{ A_{_{\infty }}(\widehat{e}_{%
\mathbf{r}},t),\frac{\partial }{\partial t}A_{\infty },\mathbf{\nabla }%
A_{\infty },\mathbf{\nabla }^{2}A_{\infty },\mathbf{f}_{\infty
}\right\} $
are allowed to be non-zero. Assumptions (\ref{minimal functional setting N-1}%
)-(\ref{minimal functional setting N-3}) are denoted as
\emph{minimal functional setting (MF setting)}.

\subsection{Scheme of presentation}

\label{sec:3b}

The plan of the paper is as follows. First (in Sec.2), the IKT
developed by Tessarotto and Ellero
\cite{Ellero2005,Tessarotto2006,Tessarotto2007} is recalled and
its basic assumptions are pointed out. Basic consequences of the
kinetic theory are analyzed in the subsequent sections 3 and 4. In
particular, in Sec.3 the Navier-Stokes dynamical system, defined
by the integral curves of the kinetic equation, is introduced and
its conditions of existence, uniqueness and regularity are
investigated. In Sec.4, the problem of existence, uniqueness and
regularity of the kinetic distribution function, for suitable
initial and boundary conditions is analyzed, with particular
reference to Maxwellian solutions. This permits us to obtain a
theorem of existence and uniqueness of strong solutions of INSE in
the MF setting.

\section{Construction of IKT for INSE}

\label{sec:4}

A remarkable aspect of fluid dynamics is related to the
construction of inverse kinetic theories (IKT) for hydrodynamic
equations in which the fluid fields are identified with suitable
moments of an appropriate kinetic probability distribution.
Recently the topic has been the subject of theoretical
investigations on the incompressible Navier-Stokes (NS) equations
(INSE) \cite{Ellero2005,Tessarotto2006,Tessarotto2007}. The
importance of the IKT-approach goes beyond the academic interest.
In fact, fluid equations represent usually a mixture of hyperbolic
and elliptic PDE's, which are extremely hard to study both
analytically and numerically. As such, their investigation
represents a challenge both for mathematical analysis and for
computational fluid dynamics. For this reason in the past
alternative approaches, based on asymptotic kinetic theories, have
been devised which permit to advance in time the fluid fields, to
be determined in terms of suitable moments of an appropriate
kinetic distribution function. These methods, which approximate
the fluid equations only in an asymptotic sense, are based on the
introduction of suitably modified (fluid) equations which permit
to advance in time the fluid fields only in an approximate sense.
In particular, typically, their modified fluid equations
actually describe weakly-compressible fluids. The discovery of IKT \cite%
{Ellero2000} provides, however, a new starting point for the
theoretical and numerical investigation of hydrodynamic equations,
since it does not require any modification of the exact fluid
equations. In particular it holds for strong solutions, and
permits to advance in time exactly the fluid fields by
means of a suitable kinetic distribution function $f(\mathbf{x,}t)$. Here $%
\mathbf{x}$ is the state vector
$\mathbf{x}=(\mathbf{r}\mathbf{,v}),$ where respectively
$\mathbf{r}$ and $\mathbf{v}$ denote the corresponding
"configuration" and "velocity" vectors, and $\Gamma $ is the
phase-space
spanned by $\mathbf{x.}$ In the following we shall assume that $\Gamma $ is a%
\emph{\ }phase-space of dimension $6$. This is achieved
introducing a phase-space classical dynamical system (here denoted
\emph{NS dynamical system})

\begin{equation}
\mathbf{x}_{o}\rightarrow \mathbf{x}(t)=T_{t,t_{o}}\mathbf{x}_{o},
\label{NS dynamical system}
\end{equation}

which uniquely advances in time the fluid fields by means of an
appropriate evolution operator $T_{t,t_{o}}$ \cite{Ellero2005}.
This is assumed to be generated by a suitably smooth vector field
$\mathbf{X}(\mathbf{x},t),$
\begin{eqnarray}
&&\left. \frac{d}{dt}\mathbf{x}=\mathbf{X}(\mathbf{x},t),\right.
\label{NS-DYN-1} \\
&&\left. \mathbf{x}(t_{o})=\mathbf{x}_{o},\right.
\label{NS-DYN-2}
\end{eqnarray}%
where $\mathbf{x=(r,v})\in \Gamma =\Omega \times
\mathbb{R}
^{3}$ is a suitable state vector, $\mathbf{r}$ and $\mathbf{v}$
denoting suitable position and velocity vectors spanning
respectively the
configuration and velocity spaces $\mathbf{\Omega }$ and $\mathbf{%
\mathbb{R}
}^{3},$ and finally the set of points $\mathbf{x}(t)$ (for $t\in
I$) defines the (phase-space) Lagrangian trajectory of the NS
dynamical system (\emph{NS Lagrangian trajectory}). Therefore,
introducing the corresponding pdf (probability distribution
function) $f(\mathbf{x,}t)\geq 0,$ it fulfills necessarily in
$\Gamma $ the differential Liouville equation
\begin{equation}
Lf(\mathbf{x},t)=0,  \label{Liouville}
\end{equation}%
where $L$ is the Liouville streaming operator
\begin{equation}
Lf\equiv \frac{\partial }{\partial t}f+\frac{\partial }{\partial \mathbf{x}}%
\cdot \left\{ \mathbf{X}(\mathbf{x},t)f\right\} .
\end{equation}%
Eq.(\ref{Liouville}) can be interpreted as a Lagrangian
\emph{inverse kinetic equation}. The corresponding equivalent
\emph{Lagrangian form} reads
\begin{equation}
J(\mathbf{x}(t),t)f(\mathbf{x}(t),t)=f(\mathbf{x}_{o},t_{o})\equiv f_{o}(%
\mathbf{x}_{o}),  \label{Eq.2}
\end{equation}%
where $f(\mathbf{x}(t),t)$ is the Lagrangian representation of the pdf, $%
\mathbf{x}(t)$ is the solution of the initial-value problem (\ref{NS-DYN-1}%
)-(\ref{NS-DYN-2}), $f_{o}(\mathbf{x}_{o})$ is{\ a suitably smooth
initial
pdf and }%
\begin{equation}
J(\mathbf{x}(t),t)=\left\vert \frac{\partial \mathbf{x}(t)}{\partial \mathbf{%
x}_{o}}\right\vert  \label{Jacobian}
\end{equation}%
is the Jacobian of the map $\mathbf{x}_{o}\rightarrow
\mathbf{x}(t).$

The vector field $\mathbf{X}(\mathbf{x},t)$ is in principle
completely arbitrary. Therefore it can be defined in such a way
that the inverse kinetic equation (\ref{Liouville}) satisfies an
appropriate set of constraint equations and in particular so that
it admits as a particular
solution the local Maxwellian distribution%
\begin{equation}
f_{M}(\mathbf{x,}t)=\frac{\rho _{o}}{\pi ^{3/2}v_{th}^{3}}\exp \left\{ -%
\frac{u^{2}}{v_{th}^{2}}\right\} ,  \label{Maxwellian}
\end{equation}%
where\emph{\ }$\mathbf{u\equiv v-V(r},t\mathbf{)}$\emph{\ }is the
relative velocity and $v_{th}(\mathbf{r},t)$\emph{\ }denotes the
thermal velocity defined in terms of the kinetic scalar
pressure\emph{\ }$p_{1}(\mathbf{r},t)$ (see below)\emph{:}
\begin{equation}
v_{th}\equiv \sqrt{\frac{2p_{1}(\mathbf{r},t)}{\rho _{o}}}.
\end{equation}

As proven in Ref.\cite{Ellero2005}, thanks to the arbitrariness in
the definition of the vector field $f(\mathbf{x,}t),$ and in the
definition of the velocity moments\ of $f,$ this permits us to
construct an inverse kinetic theory for INSE. Hence, it follows in
particular that:

\begin{enumerate}
\item For prescribed initial pdf $f_{o}(\mathbf{x}_{o})$ the time-evolved
pdf $f(\mathbf{x}(t),t),$ solution of the inverse kinetic equation (\ref%
{Liouville}), is uniquely determined by the NS dynamical system
(\ref{NS dynamical system}). Thus, if the NS dynamical system,
solution of the initial-value-problem
(\ref{NS-DYN-1})-(\ref{NS-DYN-2}), exists then necessarily
$f(\mathbf{x}(t),t)$ defined by Eq.(\ref{Eq.2}) is a solution of
the inverse kinetic equation (\ref{Liouville});

\item For a prescribed choice of the vector field $\mathbf{X}(\mathbf{x},t)$%
, Eq.(\ref{Liouville}) is an inverse kinetic equation for INSE,
i.e., the fluid equations which define INSE - namely
Eqs.(\ref{1})-(\ref{5a}) - are
provided by a suitable set of velocity-moment equations of (\ref{Liouville}%
). Therefore, if $f(\mathbf{x},t)$ is solution of the inverse
kinetic equation (\ref{Liouville}), necessarily a suitable subset
of its velocity moments coincide with the fluid fields $\left\{
\rho ,\mathbf{V},p\right\} ,$ which are solution of INSE$.$ As a
consequence, the corresponding velocity moments of
Eq.(\ref{Liouville}) must coincide identically with INSE.
\end{enumerate}

\subsection{Assumptions of IKT}

\label{sec:5}

The IKT approach for INSE and the corresponding NS dynamical
system can be
obtained in a straightforward way following the approach of Refs.\cite%
{Ellero2004,Ellero2005}. For this purpose, let us require that:

\begin{itemize}
\item The pdf $f(\mathbf{x,}t)$ is summable in velocity space, in the sense
that the velocity moments%
\begin{equation}
F_{G}(\mathbf{r},t)=\int_{V}d^{3}\mathbf{v}G(\mathbf{x,}t)f(\mathbf{x,}t)
\end{equation}%
exist in the closure domain $\overline{\Omega }\times I$ and
result suitably
smooth in $\Omega \times I$ at least for the weight functions $G(\mathbf{x,}%
t)=1,\mathbf{v,}E\equiv \frac{1}{3}u^{2},\mathbf{uu,}E\mathbf{u,}$ where $%
\mathbf{u\equiv v-V(r,}t\mathbf{)}$ denotes the relative velocity
(with respect to the fluid velocity).

\item The fluid fields $\left\{ \rho ,\mathbf{V,}%
p\right\} $ coincide with the moments:%
\begin{equation}
\left\{
\begin{array}{l}
\rho (\mathbf{r,}t)=\int_{V}d^{3}\mathbf{v}f(\mathbf{x,}t)=\rho _{o}>0, \\
\mathbf{V(r,}t\mathbf{)=}\frac{1}{\rho }\int_{V}d^{3}\mathbf{vv}f(\mathbf{x,}%
t), \\
p(\mathbf{r,}t)=p_{1}(\mathbf{r,}t)-P_{o}, \\
p_{1}(\mathbf{r,}t)=\int_{V}d^{3}\mathbf{v}\frac{1}{3}u^{2}f(\mathbf{x,}t),%
\end{array}%
\right.  \label{moment-1}
\end{equation}

being $P_{o}$ a positive constant and $p_{1}(\mathbf{r,}t)$
denoting the \emph{kinetic pressure}\textit{\ }which is defined so
that

\begin{equation}
p_{1}(\mathbf{r,}t)\geq P_{o},
\end{equation}

and hence it results $p(\mathbf{r,}t)>0$ in $\Omega \times I.$ In
addition we introduce the higher-order moments of the pdf:
\begin{equation}
\mathbf{Q=}\int_{V}d^{3}v\mathbf{u}Ef,
\end{equation}%
\begin{equation}
\underline{\underline{\mathbf{\Pi }}}=\int_{V}d^{3}v\mathbf{uu}f.
\end{equation}%
This implies that, by construction, the physical realizability conditions (%
\ref{5aa}) and (\ref{6aa}){\emph{\ }}are identically \ satisfied.
Furthermore, on can prove \cite{Ellero2005} that INSE are
satisfied by suitably selecting the vector field
$\mathbf{X(x,}t),$ i.e., by requiring that in the domain $\Omega
\times I$ the moment equations
\begin{equation}
\int_{V}d^{3}\mathbf{v}G(\mathbf{x,}t){\ L}f(\mathbf{x,}t)=0,
\end{equation}%
corresponding to the first three moments, i.e., $G(\mathbf{x,}t)=1,\mathbf{v,%
}\frac{1}{3}u^{2},$ coincide with the equations of INSE.

\item Let us impose suitable kinetic initial conditions on the pdf $f(%
\mathbf{x,}t).$ The initial conditions are manifestly of the form%
\begin{equation}
f(\mathbf{x,}t_{o})=f_{o}(\mathbf{x}),  \label{kinetic initial
condition -1}
\end{equation}%
where the initial pdf $f_{o}(\mathbf{x})$ is assumed to satisfy
the initial
conditions (\ref{initial contdion -1}),(\ref{initial condition -2}),(\ref%
{initial condition -3}) for the fluid fields, which requires%
\begin{equation}
\left\{
\begin{array}{l}
\rho _{o}=\int_{V}d^{3}\mathbf{v}f_{o}(\mathbf{x}), \\
\mathbf{V}_{o}(\mathbf{r})\mathbf{=}\frac{1}{\rho _{o}}\int_{V}d^{3}\mathbf{%
vv}f_{o}(\mathbf{x}), \\
p_{o}(\mathbf{r})=\int_{V}d^{3}\mathbf{v}\frac{1}{3}u^{2}fo(\mathbf{x}%
)-P_{o},%
\end{array}%
\right.  \label{kinetic initial condition -2}
\end{equation}%
where $\mathbf{u=v-\mathbf{V}_{o}(\mathbf{r})}$ and $P_{o}>0$ is
an arbitrary initial constant. Instead, the initial moments
\begin{equation}
\mathbf{Q}_{o}\mathbf{=}\int_{V}d^{3}v\mathbf{u}Ef_{o}(\mathbf{x}),
\label{moment-2}
\end{equation}%
\begin{equation}
\underline{\underline{\mathbf{\Pi }}}_{o}=\int_{V}d^{3}v\mathbf{uu}f_{o}(%
\mathbf{x}),  \label{moment-3}
\end{equation}%
are arbitrary, so that - for example - it is always possible to
impose that they vanish identically in $\Omega $ [by suitably
defining the initial kinetic distribution function
$f_{o}(\mathbf{x})$]$.$
\end{itemize}

\subsection{Additional assumptions - Extension of the NS Lagrangian trajectories on
$\partial \Omega $}

\label{sec:6}

An arbitrary Lagrangian trajectory defined by the NS dynamical
system can generally reach the boundary $\partial \Omega ,$ so
that the solution of the
initial value problem (\ref{NS-DYN-1})-(\ref{NS-DYN-2}) $\mathbf{x}(t)%
\mathbf{=}\left\{ \mathbf{r}(t)\mathbf{,v}(t)\right\} $ generally
may not be defined in the whole existence domain of the NS fluid
fields ($I$). This requires its extension on the boundary
$\partial \Omega.$ The result is achieved by defining suitable
\emph{boundary conditions} \emph{for the NS dynamical system,}
which must apply for an arbitrary phase-space trajectory
$\mathbf{x}(t)$ of the same dynamical system. Certain restrictions
must be placed on the possible motion of the boundary.\ This is
due both to the assumed regularity of the fluid fields [i.e., the
settings (\ref{minimal functional setting N-1}) and (\ref{minimal
functional setting N-2})] and the no-slip conditions to be imposed
on the fluid velocity due to the Dirichlet boundary conditions
(\ref{boundary condition}). In particular, we shall require that:

\begin{itemize}
\item If $\mathbf{r}_{W}$ is an arbitrary point of $\partial \Omega ,$ its
velocity, defined as $\mathbf{V}_{W}(t)\equiv \mathbf{V}_{W}(\mathbf{r}%
_{W}(t),t)=\frac{d}{dt}\mathbf{r}_{W}(t),$ is by assumption a
smooth real function of time in the sense (\ref{minimal functional
setting N-2b}).
\end{itemize}

Nevertheless, the precise nature of $\partial \Omega $ (i.e., if
it is a surface, curve or an isolated point, remains in principle
largely arbitrary. Thus, for example, $\partial \Omega $ may be
assumed as formed by piece-wise surfaces or curves, as well as
isolated
points, of $%
\mathbb{R}
^{3}$. The definition of the boundary conditions for the NS
dynamical system can, in fact, be achieved in all such these
cases.

For this purpose, let us consider an arbitrary Lagrangian trajectory $%
\mathbf{x}(t)\equiv \left\{ \mathbf{r}(t),\mathbf{v}(t)\right\} $
which at
time $t=t_{c}$ reaches the boundary $\partial \Omega $ at the position $%
\mathbf{r}_{W}(t_{c})$ with nonvanishing relative velocity, i.e.,
is such
that%
\begin{eqnarray}
&&\left. \mathbf{r}(t_{c})=\mathbf{r}_{W}(t_{c}),\right. \\
&&\left. \lim_{t\rightarrow t_{c}^{-}}\left\vert \mathbf{v}(t)-\mathbf{V}%
_{W}(t)\right\vert >0.\right.
\end{eqnarray}%
Then, introducing the unit vector $n_{w}(t_{c})$ defined so that
\begin{equation}
n_{w}(t_{c})=\lim_{t\rightarrow t_{c}^{-}}\frac{\mathbf{r}_{W}(t)-\mathbf{r}%
(t)}{\left\vert \mathbf{r}_{W}(t)-\mathbf{r}(t)\right\vert },
\end{equation}%
let us denote by $\mathbf{x}^{(-)}(t_{c})$ and
$\mathbf{x}^{(+)}(t_{c}),$
respectively, the \emph{incoming} and \emph{outgoing} Lagrangian trajectories%
$,$ which are defined as:%
\begin{equation}
\mathbf{x}^{(\pm )}(t_{c})=\lim_{t\rightarrow t_{c}^{(\pm
)}}\mathbf{x}(t).
\end{equation}%
The boundary conditions for the NS dynamical system are obtained
by imposing
the \emph{bounce-back boundary conditions }\cite{Ellero2005}%
\begin{eqnarray}
\mathbf{r}^{(+)}(t_{c}) &=&\mathbf{r}^{(-)}(t_{c}),  \label{B-B-1} \\
\mathbf{v}^{(+)}(t_{c})-\mathbf{V}_{W}(t_{c}) &=&-\left[ \mathbf{v}%
^{(-)}(t_{c})-\mathbf{V}_{W}(t_{c})-\right] .  \label{B-B-2}
\end{eqnarray}

\subsection{Kinetic boundary conditions}

\label{sec:7}

To complete the set of assumptions required by IKT, appropriate
kinetic boundary conditions must be defined for $f(\mathbf{x},t)$.
Consistent with
Eqs.(\ref{B-B-1}) and (\ref{B-B-2}), they are achieved requiring that $f(%
\mathbf{x},t)$ satisfies on $\partial \Gamma $ the following
constraints (A-C):

\begin{itemize}
\item A) kinetic \textit{bounce-back condition}: this is obtained by
imposing the conservation of probability density at $t_{c},$ i.e.,

\begin{equation}
f^{(+)}(\mathbf{r}_{W}(t_{c}),\mathbf{v}^{(+)}(t_{c})\mathbf{,}%
t_{c})=f^{(-)}(\mathbf{r}_{W}(t_{c}),2\mathbf{V}_{W}(t)-\mathbf{v}%
^{(-)}(t_{c})\mathbf{,}t_{c}),  \label{kinetic boundary condition
-1}
\end{equation}%
where $\mathbf{V}_{W}(t_{c})\equiv \mathbf{V}_{W}(\mathbf{r}%
_{W}(t_{c}),t_{c})$ denotes again the velocity of the boundary
$\partial \Omega $\ at the position $\mathbf{r}_{W}(t_{c});$

\item B) \textit{first fluid constraint}: it is provided by the requirement
that the first moment of the pdf yields the mass density, i.e., there results%
\begin{equation}
\rho
=\int_{V}d^{3}\mathbf{v}f(\mathbf{r}_{W}(t_{c})\mathbf{,v,}t_{c}).
\label{kinetic boundary condition -2}
\end{equation}

\item C) \textit{second fluid constraint}: it requires that, consistent with
the Dirichlet boundary condition for the fluid pressure [provided by Eqs.(%
\ref{boundary condition})], on $\partial \Omega $ the kinetic
pressure satisfies the constraint
\begin{equation}
p_{1}(\mathbf{r}_{W},t)-P_{o}=p_{W}(\mathbf{r}_{W},t),
\label{kinetic boundary condition -3}
\end{equation}%
where $p_{W}(\mathbf{r}_{W},t)$ and $V_{W}(\mathbf{r}_{W},t_{c})$
are
prescribed in accordance with Eqs.(\ref{boundary condition}) and $p_{1}(%
\mathbf{r}_{W},t),$ due to the position (\ref{moment-1}), reads
\begin{equation}
p_{1}(\mathbf{r}_{W},t_{c})=\int_{V}d^{3}\mathbf{v}\frac{1}{3}\left\{
\mathbf{v-V}_{W}(\mathbf{r}_{W}(t_{c}),t_{c})\right\} ^{2}f(\mathbf{r}%
_{W}(t_{c})\mathbf{,v,}t_{c}).
\end{equation}%
\emph{\ }
\end{itemize}

\subsection{Determination of the NS vector field}

\label{sec:8}

Provided the fluid fields $\left\{ \rho ,\mathbf{V,}p,\underline{\underline{%
\mathbf{\Pi }}},\mathbf{Q}\right\} $ and the volume force density $\mathbf{%
f(r,V},t)$ are continuous in $\overline{\Omega }\times I$ and
suitably
smooth, the vector field $\mathbf{X(x,}t)$ is found \cite%
{Ellero2004,Ellero2005} to be of the form $\mathbf{X(x,}t)\equiv
\left\{ \mathbf{v,F}\right\} .$ Here the vector field
$\mathbf{F\equiv F(x,}t;f)$
can be written as%
\begin{equation}
\mathbf{F(x,}t;f)=\mathbf{F}_{0}\mathbf{(x,}t;f)+\ \mathbf{F}_{1}\mathbf{(x,}%
t;f),  \label{eq. force F}
\end{equation}%
where the two vector fields $\mathbf{F}_{0}\mathbf{(x,}t;f)$ and\ $\mathbf{F}%
_{1}\mathbf{(x,}t;f),$ which depend functionally on the kinetic
distribution
$f(\mathbf{x,}t)$ (via the moments $\left\{ \rho ,\mathbf{V,}p,\underline{%
\underline{\mathbf{\Pi }}},\mathbf{Q}\right\} $)$,$ are defined
respectively as:

\begin{equation}
\mathbf{F}_{0}\mathbf{(x,}t;f)=\frac{1}{\rho }\mathbf{\nabla \cdot }%
\underline{\underline{\mathbf{\Pi }}}-\mathbf{\nabla }p_{1}\mathbf{(r,}t)+%
\mathbf{f}+\mathbf{u}\cdot \nabla \mathbf{V+}\nu \nabla
^{2}\mathbf{V,} \label{gen. force 1}
\end{equation}

\begin{equation}
\mathbf{F}_{1}\mathbf{(x,}t;f)=\frac{1}{2}\mathbf{u}\left\{ \frac{\partial }{%
\partial t}\ln p_{1}-\right.  \label{gen. force 2}
\end{equation}%
\begin{equation*}
-\frac{1}{p_{1}}\mathbf{V\cdot }\left[ \rho \frac{\partial }{\partial t}%
\mathbf{V+}\rho \mathbf{V\cdot \nabla V-f}-\mu \nabla
^{2}\mathbf{V}\right] \mathbf{+}
\end{equation*}%
\begin{equation*}
\left. +\frac{1}{p_{1}}\mathbf{\nabla \cdot
Q}-\frac{1}{2p_{1}}\left[
\mathbf{\nabla \cdot }\underline{\underline{\mathbf{\Pi }}}\right] \mathbf{%
\cdot Q}\right\} +\frac{v_{th}^{2}}{2p_{1}}\mathbf{\nabla \cdot }\underline{%
\underline{\mathbf{\Pi }}}\left\{ \frac{u^{2}}{v_{th}^{2}}-\frac{3}{2}%
\right\} .
\end{equation*}

Then the following theorem (proven in Ref.\cite{Ellero2005}) has
the flavor of: \bigskip

\textbf{THM. 1 - IKT for INSE}

\emph{Let us assume that:}

\emph{1) the fluid fields }$\left\{ \rho ,\mathbf{V,}p,\underline{\underline{%
\mathbf{\Pi }}},\mathbf{Q}\right\} $ \emph{and the volume force }$\mathbf{f}(%
\mathbf{r,}t)$ \emph{belong to the functional class defined by the
MF setting (\ref{minimal functional setting N-1})-(\ref{minimal
functional setting N-3});}

\emph{2) the pdf }$f(\mathbf{x,}t)$ \emph{is strictly positive
and} \emph{is a particular solution of IKE [Eq.(\ref{Liouville})]
in }$\Omega \times I$;

\emph{3) the velocity moments of }$f(\mathbf{x,}t)$ \emph{are
defined by Eqs.(\ref{moment-1}),(\ref{moment-2}) and
(\ref{moment-3});}

\emph{4) the pdf }$f(\mathbf{x,}t)$ \emph{satisfies the initial conditions (%
\ref{kinetic initial condition -1}) and (\ref{kinetic initial condition -2});%
}

\emph{5) the pdf }$f(\mathbf{x,}t)$ \emph{satisfies on }$\partial
\Omega $
\emph{the kinetic boundary conditions (\ref{kinetic boundary condition -1}),(%
\ref{kinetic boundary condition -2}) and (\ref{kinetic boundary condition -3}%
).}

\emph{Then it follows that}

$T1_{1})$\emph{\ }$\left\{ \rho ,\mathbf{V,}p\right\} $\emph{\ are
solutions of INSE in }$\Omega \times I;$

$T1_{2})$\emph{\ if }$f(\mathbf{x},t)$\emph{\ satisfies in the
whole
phase-space }$\Gamma =\Omega \times I$\emph{\ the initial conditions }$f(%
\mathbf{x},t_{o})=f_{M}(\mathbf{x},t_{o}),$\emph{\ and on }$\Omega
\delta
\times I$\emph{\ the boundary conditions }%
\begin{equation}
f^{(+)}(\mathbf{r}_{W}(t_{c}),\mathbf{v,}t)=f^{(-)}(\mathbf{r}_{W},2\mathbf{V%
}_{W}(t)-\mathbf{v,}t),
\end{equation}%
\emph{then it follows that }$f(\mathbf{x},t)$\emph{\ is solution of IKE [Eq.(%
\ref{Liouville})] in the whole set }$\Gamma =\Omega \times I.$

$T1_{3})$\emph{\ if }$f(\mathbf{x},t)$\emph{\ is a particular
solution of IKE, then }$\left\{ \rho ,\mathbf{V,}p\right\}
$\emph{\ are solutions of INSE in }$\Omega \times I.$

\section{The Navier-Stokes dynamical system: a theorem of local existence
and uniqueness}

\label{sec:9}

Let us now analyze the existence, uniqueness and regularity of the
IKT. In the following sections we intend to address in particular
the well-posedness of the initial-value problem
(\ref{NS-DYN-1})-(\ref{NS-DYN-2}) which defines the Navier-Stokes
dynamical system$.$

It is immediate to prove that under suitable smoothness assumptions on $%
\left\{ \rho ,\mathbf{V,}p,\underline{\underline{\mathbf{\Pi }}},\mathbf{Q}%
\right\} _{(\mathbf{r,}t)}$, the problem
(\ref{NS-DYN-1})-(\ref{NS-DYN-2}) admits a local existence and
uniqueness theorem and therefore defines a dynamical system
$S_{NS},$ to be denoted as \textit{Navier-Stokes dynamical system.
} \bigskip

\textbf{THM. 2 - Local existence, uniqueness and regularity of the
NS dynamical system}

\emph{Let us require that the fluid fields }$\left\{ \rho ,\mathbf{V,}p,%
\underline{\underline{\mathbf{\Pi }}},\mathbf{Q}\right\} $
\emph{and the
volume force density} $\mathbf{f(r,}t)$\emph{\ are bounded in }$\overline{%
\Omega }\times I$ \emph{and belong to the MF setting (\ref{minimal
functional setting N-1})-(\ref{minimal functional setting N-3}).}

\emph{Then it follows that:}

\emph{a) the vector field }$\mathbf{X(x,}t)=\left[
\mathbf{v,F(x,}t)\right] $ \emph{is of class }$C^{(1,1)}(\Gamma
\times I);$\emph{\ }

\emph{b) the solution of the problem
(\ref{NS-DYN-1})-(\ref{NS-DYN-2})
exists locally and is unique: }%
\begin{equation*}
\mathbf{x(}t\mathbf{)=\chi (x}_{o},t_{o},t)\equiv \left\{ \mathbf{r(}t%
\mathbf{),v(}t\mathbf{)}\right\} \equiv \left\{ \mathbf{\chi }_{\mathbf{r}}%
\mathbf{(x}_{o},t_{o},t),\mathbf{\chi }_{\mathbf{v}}\mathbf{(x}%
_{o},t_{o},t)\right\} .
\end{equation*}%
\emph{Moreover:}

\emph{c) }$\mathbf{x(}t\mathbf{)=\chi (x}_{o},t_{o},t)$ \emph{and
its inverse,} $\mathbf{x}_{o}\mathbf{=\chi (x}(t),t,t_{o})$\emph{\
are at least:}
\begin{equation}
\mathbf{\chi (x}_{o},t_{o},t),\mathbf{\chi (x}(t),t,t_{o})\in
C^{(2)}(\Gamma \times I\times I);  \label{reg v}
\end{equation}

\emph{d) the Jacobian} $J(\mathbf{x(}t\mathbf{),}t)$ \emph{of the
flow}
\begin{equation}
\mathbf{x}_{o}\rightarrow \mathbf{x(}t)=\chi
(\mathbf{x}_{o},t_{o},t) \label{N-S phase-flow}
\end{equation}%
\emph{is at least of class}
\begin{equation}
J(\mathbf{x(}t\mathbf{),}t)\in C^{(1)}\left( \Gamma \times
I\right) \label{reg J}
\end{equation}%
\emph{and for all} $\mathbf{x}_{o}\in \Gamma $ \emph{and} \emph{for all} $%
t_{o},t\in I$\emph{\ is non-vanishing and finite, i.e., }%
\begin{equation}
J(\mathbf{x(}t\mathbf{),}t)\neq 0,\infty .  \label{value of J}
\end{equation}%
\qquad

\textit{Proof - }In validity of the regularity assumption the vector field $%
\mathbf{X(x,}t\mathbf{)}$ is manifestly of class $C^{(1,1)}(\Gamma
\times I). $ Hence, thanks to the fundamental theorem of existence
and uniqueness (see for example, Coddington and Levinson, 1955
\cite{Coddington1955},
Hirsch and Smale, 1974 \cite{Hirsch1975}) the regularity conditions (\ref%
{reg v}) are implied. The additional result (\ref{reg J}) follows
directly from Eq.(\ref{Jacobian}) and the definition given above
for the vector field $\mathbf{F(x,}t)$ [Eq.(\ref{eq. force F})].
Moreover, thanks to Liouville theorem the Jacobian
$J(\mathbf{x(}t\mathbf{),}t)$ does not vanish or diverge for all
$\mathbf{x}_{o}\in \Gamma $ and for all $t_{o},t\in I.$ Q.E.D.
\bigskip

An immediate consequence of THM.2 is the following \bigskip

\textbf{COROLLARY 1 of THM.2 - Minimal functional setting of the fluid fields%
}

\emph{If the fluid fields obey the MF setting (\ref{minimal
functional setting N-1})-(\ref{minimal functional setting N-3})
there follows in particular that:}

\emph{a) the NS vector field }$\mathbf{X(x,}t)=\left[ \mathbf{v,F(x,}t)%
\right] $ \emph{results of class }$C^{(1,\infty ,1)}(\Omega \times
V\times I);$\emph{\ }

\emph{b) the problem (\ref{NS-DYN-1})-(\ref{NS-DYN-2}) admits one
and only one solution }$\mathbf{x(}t)$ \emph{of class} $%
C^{(1,2,2)}(\Gamma \times I\times I);$

\emph{c) the Jacobian} $J(\mathbf{x(}t\mathbf{),}t)$ \emph{of the
phase-flow is at least} $C^{(1,2)}\left( \Gamma \times I\right)$.

\textit{Proof \ }  - The result follows from THM's 1 and 2 and the
assumption of the MF setting for the fluid fields. Q.E.D.
\bigskip

From THM.2 there follows the obvious further corollary:

\bigskip

\textbf{COROLLARY 2 of THM.2 } - \textbf{Extension of the solution} $\mathbf{%
x(}t)$\textbf{\ on} $\partial \Omega $

\emph{The solution} $\mathbf{x(}t\mathbf{)=\chi (x}_{o},t_{o},t)$
\emph{of the initial value problem
Eq.(\ref{NS-DYN-1})-(\ref{NS-DYN-2}), prolonged on}
$\partial \Omega $ \emph{by means of the bounce-back boundary condition (\ref%
{kinetic boundary condition -1})}, \emph{exists and is unique. }

\textit{Proof }- The result is an immediate consequence of THM.1,
the assumption of validity of the MF setting and the boundary
conditions previously introduced (see Sec.2.3) for the NS
dynamical system. Q.E.D.\

\section{The initial-boundary value problem for the kinetic pdf and INSE:
existence, uniqueness and regularity}

\label{sec:10} The previous theorem of local existence and
uniqueness for the NS dynamical system, and its extension given by
the 2nd Corollary of THM.2, can now be used to obtain a local
theorem of existence, uniqueness and regularity for the kinetic
distribution function $f(\mathbf{x,}t).$ Due to its arbitrariness,
it is always possible to limit ourselves to the investigation of
initial conditions of the form
\begin{equation}
f(\mathbf{x,}t_{o})=f_{M}(\mathbf{x,}t_{o}).  \label{choice}
\end{equation}%
In view of the positions (\ref{eq. force F}),(\ref{gen. force
1})(\ref{gen.
force 2}) one can prove (see below) that this implies identically $f(\mathbf{%
x,}t)\equiv f_{M}(\mathbf{x,}t).$ We stress that a more general
result holds
for an arbitrary (but suitably smooth and summable) distribution $f(\mathbf{%
x,}t)$ (see related discussion in Ref.\cite{Ellero2005}). \ The choice (\ref%
{choice}) warrants summability and existence of all the required
velocity moments and in addition is consistent with the physical
requirement set by
the PEM (principle of entropy maximization; see Refs.\cite%
{Tessarotto2008a,Tessarotto2008b}). Then, the following theorem
holds:
\bigskip

\textbf{THM. 3 - Local existence and uniqueness of the kinetic
pdf}

\emph{Let us require that:}

\emph{1) the initial kinetic pdf} $f_{o}(\mathbf{x})$\emph{\
coincides with the Maxwellian kinetic pdf
}$f_{M}(\mathbf{x,}t_{o})$\emph{;}

\emph{2) the fluid fields} $\left\{ \rho ,\mathbf{V,}p\right\} _{(\mathbf{r}%
,t)}$ \emph{and the volume force density }$\mathbf{f(r,}t)$\emph{\
belong to
the MF setting defined by Eqs.(\ref{minimal functional setting N-1})-(\ref%
{minimal functional setting N-3});}

\emph{3) the fluid fields satisfy the fluid initial and boundary conditions (%
\ref{initial contdion -1})-(\ref{initial condition -3}) and
(\ref{boundary condition});}

\emph{4) for all }$\left( \mathbf{x,}t\right) \in \overline{\Gamma
}\times I$ \emph{the kinetic pdf }$f(\mathbf{x},t),$\emph{\ if it
exists, satisfies the bounce-back condition and the conditions on
pressure (\ref{kinetic boundary condition -1}),(\ref{kinetic
boundary condition -2}) and (\ref{kinetic boundary condition
-3});}

\emph{It follows that in the domain }$\Gamma \times I:$\emph{\ }

\emph{a)} $f_{M}(\mathbf{x,}t)$ \emph{is a particular solution of IKE [Eq.(%
\ref{Liouville})];}

\emph{b)} $f_{M}(\mathbf{x,}t)$ \emph{is differentiable;}

\emph{c)} $f_{M}(\mathbf{x,}t)$ \emph{is summable; }

\emph{d) the velocity moments of }$f_{M}(\mathbf{x,}t),$
\emph{corresponding to} $G(\mathbf{x,}t)=1,\mathbf{v,}E\equiv
\frac{1}{3}u^{2}$, \emph{coincide with the fluid fields }
\begin{equation}
\left\{\rho ,\mathbf{V,}p_{1}\equiv p+P_{o}\right\}
_{(\mathbf{r},t)};
\end{equation}

\noindent \emph{and moreover:}

\emph{e) }$f_{M}(\mathbf{x,}t)$ \emph{is defined and summable also
on the boundary set }$\partial \Gamma \times I,$ \emph{where
}$\partial \Gamma =\partial \Omega \times V.$

\textit{Proof - }The proof of a) follows by direct substitution of
the position $f(\mathbf{x,}t)=f_{M}(\mathbf{x,}t)$ in
Eq.(\ref{Liouville}). In addition, Eq.(\ref{Eq.2}), shows that
this solution necessarily corresponds only to the initial
condition (\ref{choice}), i.e., there results necessarily in
$\Gamma \times I:$

\begin{equation}
f_{M}(\mathbf{x}(t),t)=\frac{f_{M}(\mathbf{x}_{o},t_{o})}{J(\mathbf{x}(t),t)}%
.
\end{equation}

b) Thanks to assumptions 1)-4), $f_{M}(\mathbf{x,}t)$ is
manifestly
differentiable in $\overline{\Gamma }\times I$ . c) Similarly, $f_{M}(%
\mathbf{x,}t)$ is manifestly summable and d) its moments corresponding to $G(%
\mathbf{x,}t)=1,\mathbf{v,}E\equiv \frac{1}{3}u^{2}$ are by definition $%
\left\{ \rho ,\mathbf{V,}p_{1}\equiv p+P_{o}\right\}
_{(\mathbf{r},t)}$. d) Finally, thanks to Corollary 2 of THM.2,
$f_{M}(\mathbf{x,}t)$ is also manifestly defined on the boundary
$\partial \Omega .$ Q.E.D.

THM.3 already contains in itself the basic ingredients required to
reach the theorem of existence and uniqueness for INSE, which can
be written as: \bigskip

\textbf{THM. 4 - Local existence and uniqueness of the NS fluid
fields}

\emph{In validity of THM.3 it follows that in the domain }$\Omega \times I,$%
\emph{\ the fluid fields }$\left\{ \rho ,\mathbf{V,}p\right\} _{(\mathbf{r}%
,t)}$\emph{\ are necessarily strong solutions of the
initial-boundary value problem of INSE.}

\textit{Proof - }In fact, thanks to THM.1 and 3 it follows that $f_{M}(%
\mathbf{x,}t)$ is a particular solution of IKE
[Eq.(\ref{Liouville})] if and only if\ $\left\{ \rho
,\mathbf{V,}p\right\} _{(\mathbf{r},t)}$\emph{\ }are necessarily
strong solutions of the initial-boundary value problem of INSE
which belong to the MF setting. As a consequence the moments of $f_{M}(%
\mathbf{x,}t),$ $\left\{ \rho ,\mathbf{V,}p_{1}\right\}
_{(\mathbf{r},t)}$ obey, by construction, INSE and the associated
initial and boundary conditions, set by prescribing
$f_{M}(\mathbf{x,}t_{o})$ and respectively{} Eqs.(\ref{kinetic
boundary condition -1}),(\ref{kinetic boundary condition -2}) and
(\ref{kinetic boundary condition -3}). Q.E.D.

\bigskip

We remark that, in the case $\Omega $\ is an unbounded domain,
thanks to assumptions (\ref{minimal functional setting N-3}), in
principle no
restriction is placed on the asymptotic behavior of the fluid fields for $%
\left\vert \mathbf{r}\right\vert \rightarrow \infty $ at time $t.$
Hence, in
contrast to the customary approach \cite%
{Ladyzhenskaya1969,Lions,Kato1964,Temam1983}, these solutions do
not necessarily belong to Sobolev spaces.

THM's. 3 and 4 contain the main contribution of the paper and the
basic new results regarding the existence, uniqueness and
regularity of nondecaying
strong solutions of INSE. Thus, provided the fluid fields $\left\{ \rho ,%
\mathbf{V,}p\right\} _{(\mathbf{r},t)}$ satisfy the assumptions of
the MF setting in the set $\Omega \times I,$ together with the
initial and boundary
conditions, defined respectively by Eqs.(\ref{initial contdion -1})-(\ref%
{initial condition -3}) and (\ref{boundary condition}), the
solution of INSE exists and is unique.

It is possible to show that the regularity assumptions on the
fluid pressure{} are minimal in the context of the present
approach. For example, if the fluid pressure is assumed to be
unbounded from below, the inverse kinetic approach manifestly
fails, since it cannot be related to the kinetic pressure in this
case. In fact, the Maxwellian distribution is defined and results
summable in velocity space (in the sense indicated above) only if
the kinetic pressure is strictly positive $p_{1}(\mathbf{r},t)>0.$
Moreover, if\ the partial time derivative $\frac{\partial
p}{\partial t}$ is assumed unbounded or discontinuous, the NS
dynamical system cannot be defined any more, since the same NS
vector field ceases to be at least continuous.

The present result generalizes also the treatment of \ uniqueness
of nondecaying strong solutions recently given by J. Kato
\cite{Jun-Kato2003}. The following remark is relevant in this
context. First, there is a one-to-one correspondence between fluid
fields and the Maxwellian
distribution $f_{M}(\mathbf{x},t),$ while the uniqueness of $f_{M}(\mathbf{x}%
,t)$ and of the phase-flow (\ref{moment-3}) imply each other. Both
are
determined, in turn, by the solution of the initial-value problem (\ref%
{NS-DYN-1})-(\ref{NS-DYN-2}) and therefore by the value of the Jacobian $J(%
\mathbf{x}(t),t).$ Therefore, the uniqueness of strong solutions
is intrinsically related to their existence. In fact,
$J(\mathbf{x}(t),t)$ is defined and finite if and only if the
fluid fields belong to the MF setting, and in particular the
kinetic pressure $p_{1}(\mathbf{r},t)$ results strictly positive,
while - at the same time - the fluid fields are finite,
i.e., \ $p_{1}(\mathbf{r},t)<\infty $ and $\left\vert \mathbf{V}(\mathbf{r}%
,t)\right\vert <\infty .$

\section{Discussion and Conclusions}

\label{sec:11}

In this paper the problem of existence of strong solutions of the
initial-boundary value problem of INSE has been addressed in the
context of
the IKT earlier developed \cite%
{Ellero2000,Ellero2004,Ellero2005,Tessarotto2006,Tessarotto2007}.
The striking new features of present approach is that the fluid
fields, and in particular the fluid velocity, do not necessarily
decay at infinity in $\Omega ,$ hence the solution is not embedded
in Sobolev spaces. \

The proof of existence, uniqueness and regularity of strong
solutions of INSE is reached in two steps. The first one is based
on the introduction of the NS vector fields
$\mathbf{X(x,}t\mathbf{),}$ suitably related to the fluid fields
$\left\{ \rho ,\mathbf{V,}p\right\} ,$ and by establishing the
conditions of existence of its associated dynamical system, the NS
dynamical system. The second step consists in the introduction of
a suitable inverse kinetic equation and a related initial-boundary
value problem, for which a theorem of existence, uniqueness and
regularity can be reached.

As a consequence, INSE are found to be expressed in terms of
suitable velocity moments of the kinetic equation, whereas the
initial and boundary conditions for the fluid fields are satisfied
identically by proper definition of the initial and boundary
conditions for the kinetic distribution function.

A discussion on consequences and applications of IKT in fluid
dynamics can
be found in Refs.\cite{Tessarotto2008a,Tessarotto2008b} and \cite%
{Tessarotto2008t}.

Here we mention that by means of the IKT here adopted, the fluid pressure $p%
\mathbf{(r,}t\mathbf{)}$ can be advanced in time self-consistently
without solving explicitly the Poisson equation, since it can be
determined directly as a moment of the kinetic distribution
function by advancing in time the initial kinetic distribution
function $f(\mathbf{x},t_{o})$ in terms of NS dynamical system
(\ref{NS dynamical system}) and IKE (\ref{Liouville}). As a
fundamental result, an exact pressure evolution equation,
advancing in time
self-consistently the fluid pressure, can actually be achieved \cite%
{Tessarotto2008p}.

This feature is potentially relevant for the development of
numerical solution methods for INSE based on the discretization of
the kinetic distribution in phase space and yields a possible
alternative to direct solution methods based on the solution of
Poisson equation for the fluid
pressure $p$ \cite{Ellero2005} (see also for example, \cite%
{vanDoormal1984,Kim-Moin1985,Issa1986,Choi1994}).

\bigskip

\section*{Acknowledgments}
Work developed in cooperation with the CMFD Team, Consortium for
Magneto-fluid-dynamics (Trieste University, Trieste, Italy).\
Research developed in the framework of the MIUR (Italian Ministry
of University and Research) PRIN Programme: \textit{Modelli della
teoria cinetica matematica nello studio dei sistemi complessi
nelle scienze applicate} and the CMFD Consortium, University of
Trieste, Italy.


\end{document}